\documentclass[12pt,preprint]{aastex}

\def\specchar#1{{\sc #1}}
\def\FeI{\mbox{Fe\,\specchar{i}}}

\def\eg{e.g.}

\def\arcsec{\hbox{$^{\prime\prime}$}}

\def\pun{\stackrel{}{\mbox{.}}}
\def\farcs{$\stackrel{\prime\prime}{\pun}$}

\def\kms{\hbox{km$\;$s$^{-1}$}}
\def\ms{\hbox{m$\;$s$^{-1}$}}

\shorttitle{Where the granular flows bend}

\shortauthors{Khomenko et al.}

\begin{document}

\title{Where the granular flows bend}

\author{E. Khomenko\altaffilmark{1,2}, V. Mart{\'{\i}}nez Pillet \altaffilmark{1},
S.K. Solanki\altaffilmark{3,8},
J.C. del Toro Iniesta\altaffilmark{4},
A. Gandorfer\altaffilmark{3},
J.A. Bonet\altaffilmark{1},
V. Domingo\altaffilmark{5},
W. Schmidt\altaffilmark{6},
P. Barthol\altaffilmark{3},
M. Kn{\"o}lker\altaffilmark{7}}
\email{khomenko@iac.es}
\altaffiltext{1}{Instituto de Astrof\'{\i}sica de Canarias, 38205,
C/ V\'{\i}a L{\'a}ctea, s/n, La Laguna, Tenerife, Spain}
\altaffiltext{2}{Main Astronomical Observatory, NAS, 03680, Kyiv,
Ukraine}
\altaffiltext{3}{Max-Planck-Institut f\"ur
Sonnensystemforschung, 37191, Katlenburg-Lindau, Germany}
\altaffiltext{4}{Instituto de Astrof\'\i sica de Andaluc\'\i a
(CSIC), Apdo. de Correos 3004, E-18080, Granada, Spain}
\altaffiltext{5}{Grupo de Astronom\'\i a y Ciencias del Espacio
(Univ. de Valencia), E-46980, Paterna, Valencia, Spain}
\altaffiltext{6}{Kiepenheuer-Institut f\"ur Sonnenphysik, 79104,
Freiburg, Germany}
\altaffiltext{7}{High Altitude Observatory (NCAR), 80307-3000,
Boulder, USA; The National Center for Atmospheric Research is
sponsored by the National Science Foundation.}
\altaffiltext{8}{School of Space Research, Kyung Hee University,
Yongin, Gyeonggi, 446-701 Korea}

\begin{abstract}
Based on IMaX/{\sc Sunrise} data, we report on a previously
undetected phenomenon in solar granulation. We show that in a very
narrow region separating granules and intergranular lanes the
spectral line width of the \FeI\ 5250.2 \AA\ line becomes
extremely small. We offer an explanation of this observation with
the help of magneto-convection simulations. These regions with
extremely small line widths correspond to the places where the
granular flows bend from mainly upflow in granules to downflow in
intergranular lanes. We show that the resolution and image
stability achieved by IMaX/{\sc Sunrise} are important requisites
to detect this interesting phenomenon.
\end{abstract}


\section{Introduction}

Solar granulation is one of the most studied phenomena in the
lower solar atmosphere and its properties are thought to be well
understood \citep{Bray1984}. Spectral and imaging observational
data have allowed solar granulation to be studied in detail.
Commonly considered parameters include sizes and lifetimes of
granules, velocity and intensity fluctuations together with their
correlations, proper motions of individual features, etc.
\citep[e.g.,][]{Title+others1989, Karpinsky1990, Nesis+etal1993,
Espagnet+Muller+Roudier+Mein+Mein1995,
Hirzberger+Vazquez+others1997, Hirzberger1999, Kostik+etal2009}.
At the same time, state of the art simulations of solar convection
and magneto-convection reproduce very satisfactorily many
observational granulation properties, which has made it possible
to explain self-consistently the mean shapes of spectral line
profiles \citep{Nordlund+etal2009}. Simulations have shown that
parameters as micro- and macroturbulent velocities (used widely in
the past to match the spectral lines shapes to observations) are
not necessary any more as the spectral lines are broadened by the
velocity field of the convection motions.

In this Letter, we report on spatially resolved observations of
the line broadening by the granular velocity field (or,
equivalently, the spatially resolved action of solar
microturbulence). The data used in the Letter were obtained with
the Imaging Magnetograph eXperiment \citep[IMaX;
see][]{MartinezPillet+etal2010} that flew on board the {\sc
Sunrise} balloon telescope in June, 2009 \citep{Barthol+etal2010,
Solanki+etal2010}. We study the spatial variations of the full
width at half maximum (FWHM) of the \FeI\ 5250.2 \AA\ line. We
found that the FWHM is larger in intergranular lanes and smaller
in granules. But the most interesting finding is that the FWHM is
particularly small in narrow lanes located at
granular/intergranular borders.
We found only few works that discuss the relation between the line
FWHM and granulation \citep[\eg ][]{Nesis+etal1992,
Hanslmeier+etal2008}, pointing out that the FWHM is
anti-correlated with brightness. The observations of
\citet{Nesis+etal1992} were interpreted in terms of supersonic
flows, which was questioned by
\citet{Solanki+Rueedi+Bianda+Steffen1996}. We did not find any
previous mention of the detection of the narrow regions with
extremely small FWHM adjacent to intergranular borders.
The aim of this Letter is to describe their properties and to
offer an explanation for their origin.

\begin{figure*}
\center
\includegraphics[width=16cm]{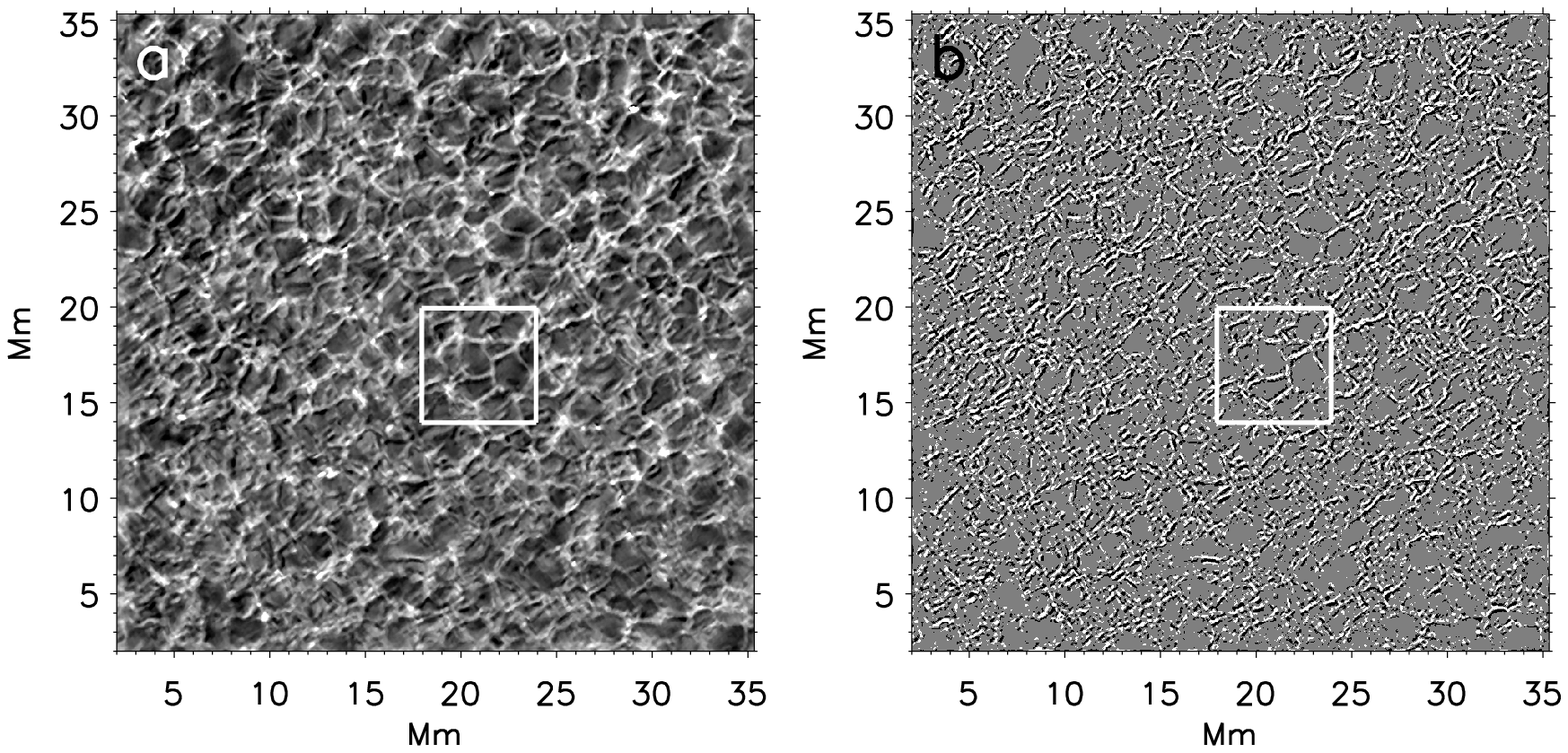}
\includegraphics[height=8cm]{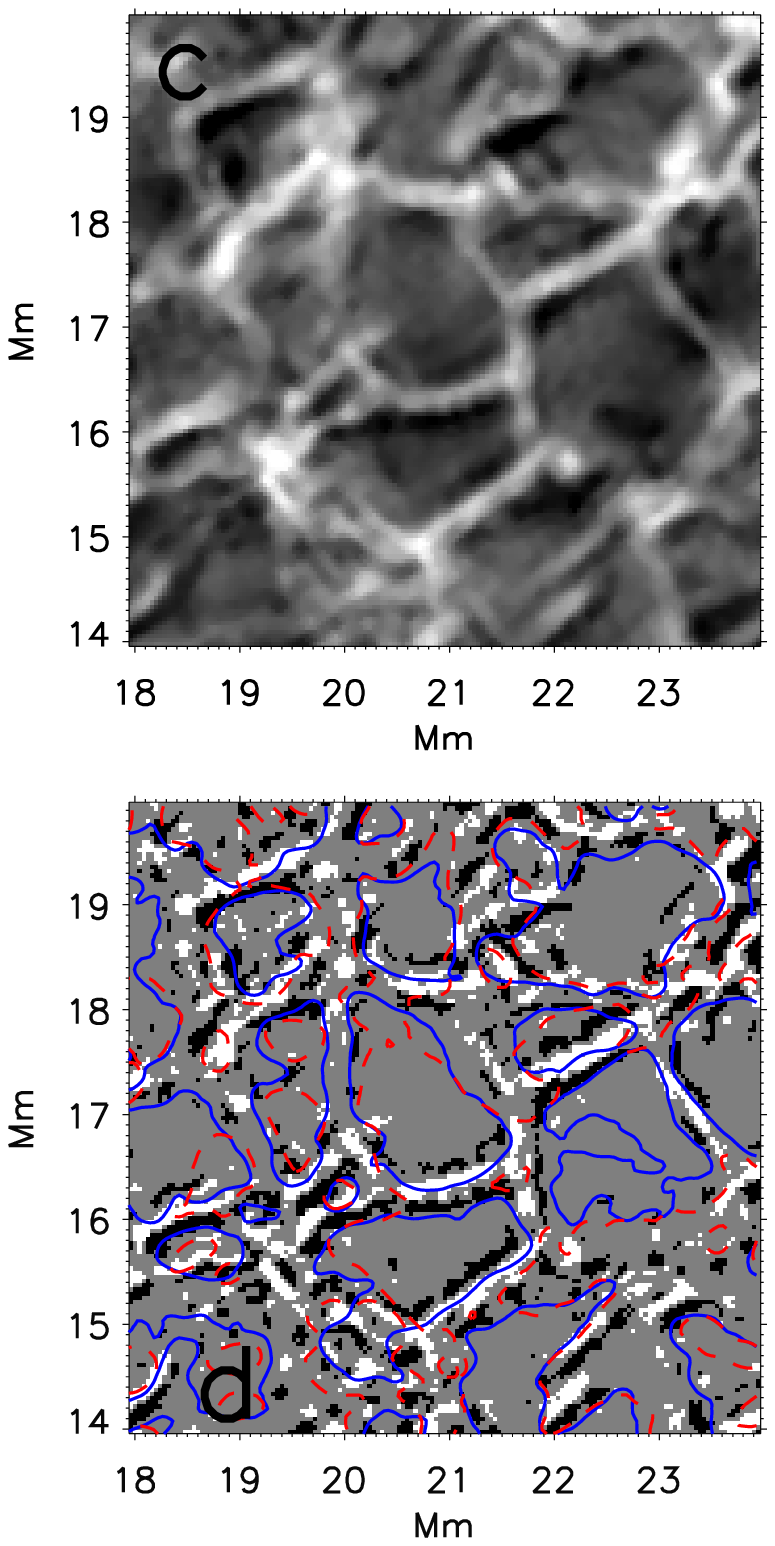}
\caption{(a) Image of the full width at half maximum (FWHM) of the
\FeI\ 5250.2 \AA\ line in the IMaX field of view; the grey scaling
ranges from 90 to 150 m\AA; the mean FWHM is equal to 115 m\AA;
(b) same image after application of the unsharp masking filter.
Panels (c) and (d) give an enlarged view of the area of 6 Mm
$\times$ 6 Mm marked by a white square in panels (a) and (b). Blue
contours in panel (d) correspond to normalized continuum intensity
$I_c /\bar{I_c}=1$. Red dashed contours mark locations with
zero line of sight velocity. }\label{fig:fwhm} \vspace{-0.5cm}
\end{figure*}

\begin{figure*}
\center
\includegraphics[width=8cm]{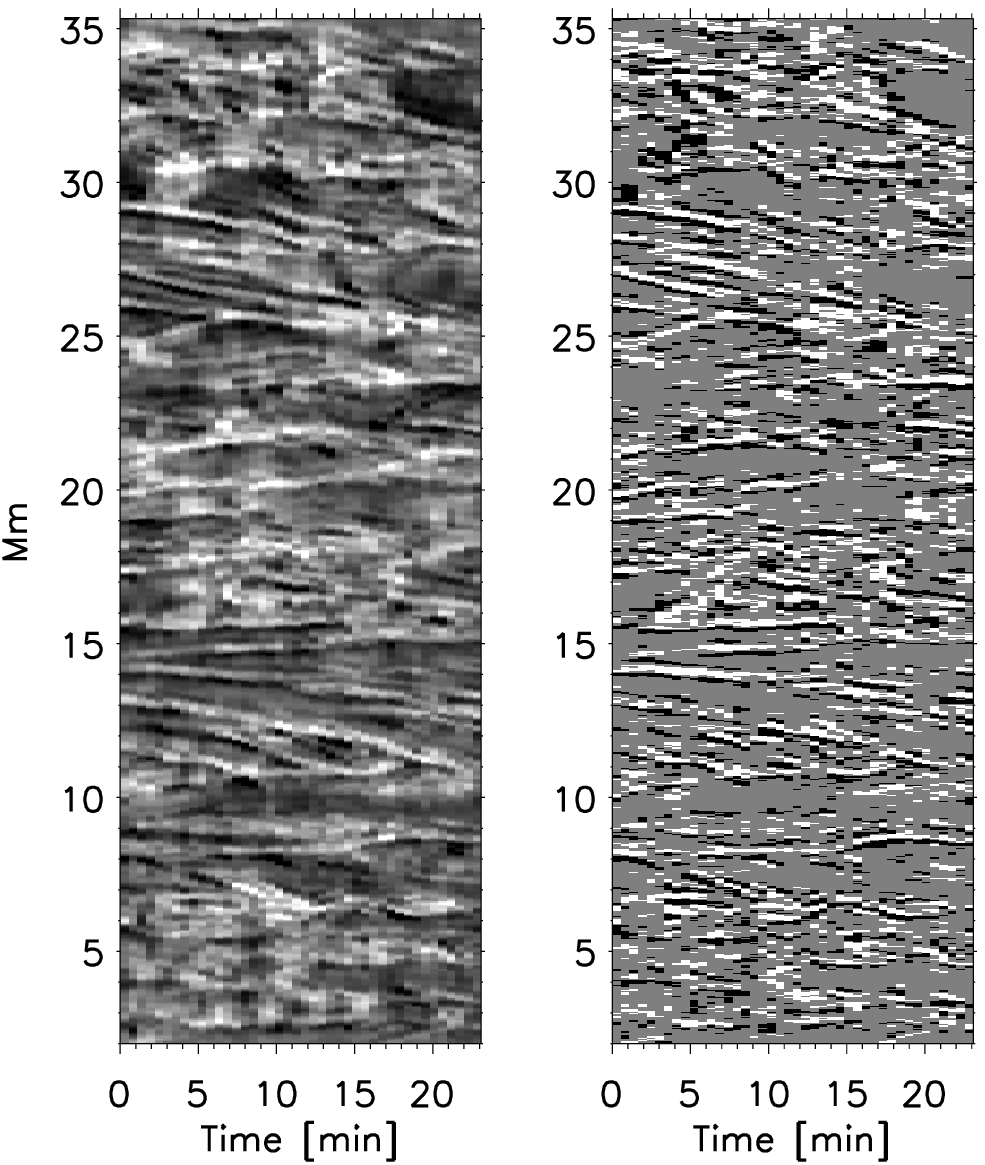}
\includegraphics[width=8cm]{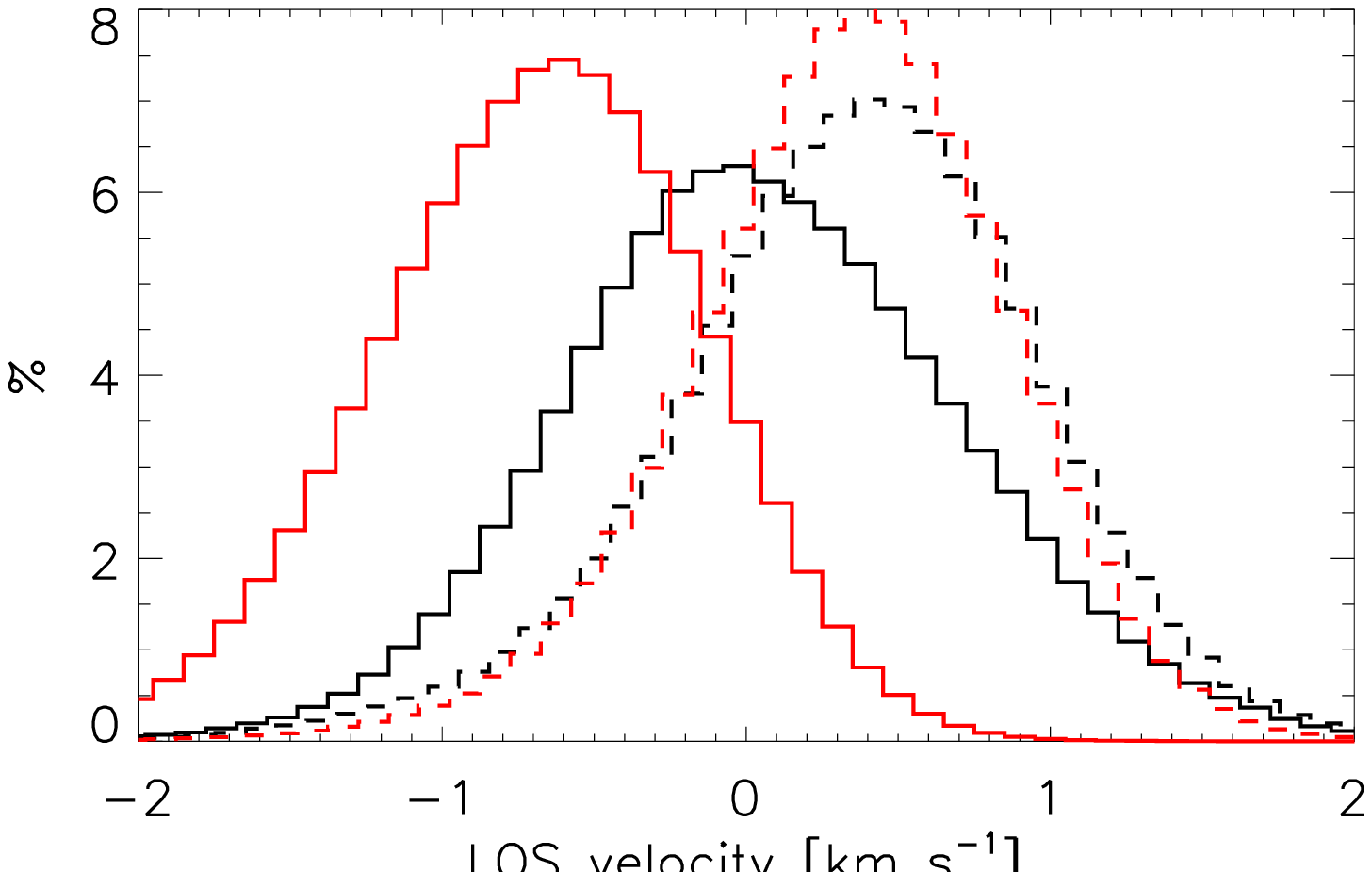}
\caption{{\it Left panel:} temporal variations of the FWHM in the
IMaX data along a virtual slit placed at one selected horizontal
location; the grey scaling ranges from 90 to 150 m\AA; the
mean FWHM is equal to 115 m\AA.  {\it Middle panel} gives the
same image after application of the unsharp masking filter. {\it
Right panel:} histograms of the LOS velocity in the IMaX data at
the locations where the continuum intensity is above unity, i.e.
in granules (red solid line); at the locations where the continuum
intensity is below unity, i.e. in intergranular lanes (red dashed
line); at the locations with enhanced FWHM marked white in
Fig.~\ref{fig:fwhm}b (black dashed line); and at locations with
extremely small FWHM marked black in Fig.~\ref{fig:fwhm}b (black
solid line).}\label{fig:time} \vspace{-0.5cm}
\end{figure*}

\section{Observed line width}

The IMaX instrument and the details of the data acquisition and
reduction are described in \citet{MartinezPillet+etal2010}. Here
we use a time series of images of about 23 min duration taken near
the solar disc center (UT 00:37). The observational sequence
consists of five polarized filtergrams taken at $\pm40$, $\pm80$
and $+227$ m\AA\ from the \FeI\ 5250.2 \AA\ line center. We use
Level 2 data (reconstructed by deconvolution). The spatial
resolution of the reconstructed data is estimated to be 0\farcs15.
The field of view is about 33\arcsec $\times$ 33\arcsec\ (after
removing the apodized areas), with a pixel size of 0\farcs055. We
use the following parameters extracted from the Stokes $I$ data:
continuum intensity, line of sight velocity, and FWHM. The latter
two parameters were calculated from Gaussian fits to the five
intensity wavelength points. The statistical errors of the
fitting procedure were within the range of 20--40 \ms\ for the
velocities and a few m\AA\ to 10 m\AA\ (at most) for the FWHM.

Figure \ref{fig:fwhm} shows an image of the FWHM at one moment of
time. This image demonstrates clearly that the width of the \FeI\
5250.2 \AA~intensity profile follows the granulation structure.
The FWHM increases in intergranular lanes (appearing as bright
regions) and becomes smaller in granules. A more careful
inspection of the image reveals that adjacent to almost each
intergranular lane there are narrow black regions (that also have
elongated lane-like shapes) of smaller widths.
The average FWHM of the \FeI\ 5250.2 line in the granular
regions is of 120 m\AA, in intergranular regions is of 140 m\AA,
but only 100 m\AA\ at the borders between granules and
intergranular lanes.
The unsharp mask filter\footnote{The algorithm that works by
enhancing the contrast between neighboring pixels in an image}
applied to this image enhances the gradients and makes the
locations with extremely small FWHM more evident
(Fig.~\ref{fig:fwhm}b).
Panels c and d with enlarged view of the region of 6 Mm $\times$ 6
Mm size (marked by the white squares in Figs.~\ref{fig:fwhm}a and
b) give some very clear examples of the dark lanes surrounding
bright intergranular spaces. These dark features sometimes appear
only on one side of the bright intergranular lane, and sometimes
on both sides. In many cases they appear in regions where the
normalized continuum intensity $I/\bar{I_c} \approx 1$, i.e. just
at the borders between granules and intergranular lanes. The width
of these dark features is about $3-4$ IMaX pixels, i.e.
$\sim$0\farcs2. Their lifetime is the same as the lifetime of the
intergranular lane they are attached to, see Fig.~\ref{fig:time}
(left and middle panels).
These dark lanes in the FWHM image appear also in the
non-reconstructed data (with less contrast), proving that they are
not artifacts of the phase diversity reconstruction procedure.

\begin{figure*}
\center
\includegraphics[width=17cm]{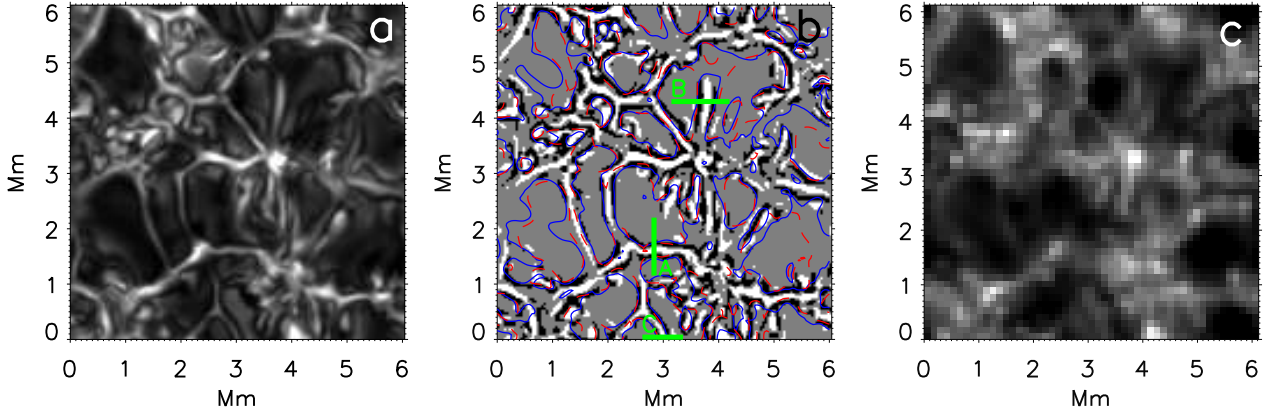}
\caption{(a) Image of the FWHM of the \FeI\ 5250.2 \AA\ line
obtained from simulations degraded to IMaX resolution, the grey
scaling ranges from 90 to 150 m\AA; the mean FWHM is equal to 112
m\AA; (b) same image after application of the unsharp masking
filter; blue contours correspond to $I_c /\bar{I_c}=1$; red
dashed contours correspond to zero line of sight velocity; green
lines indicate locations considered in Fig. 5; (c) image of the
FWHM of the \FeI\ 6301.5 \AA\ line obtained from simulations
degraded to Hinode resolution; the grey scaling ranges from
123 to 160 m\AA; the mean FWHM is equal to 132 m\AA. 
}\label{fig:fwhm_sims} \vspace{-0.5cm}
\end{figure*}

\begin{figure}
\center
\includegraphics[width=11cm]{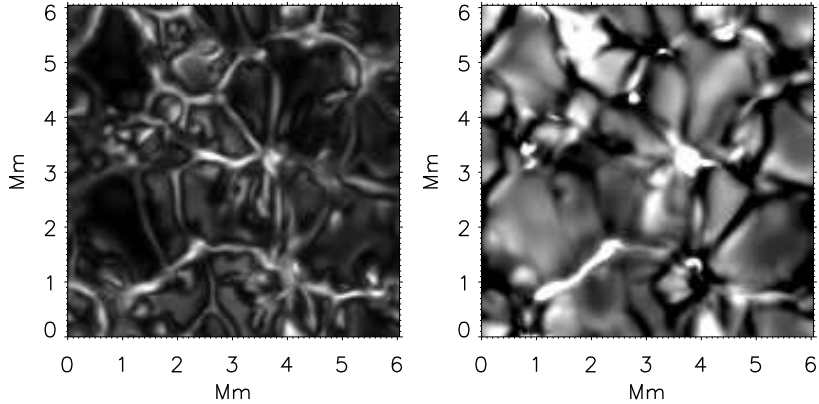}
\caption{Left panel: image of the FWHM obtained from simulations
when the spectral synthesis is done after setting horizontal
variations of thermodynamics variables to zero; the grey scaling
ranges from 90 to 150 m\AA; the average value is 112 m\AA. Right
panel: the same for the case that the synthesis is performed after
setting velocities to zero in the simulation snapshot; the grey
scaling ranges from 98 to 105 m\AA, the average value is 100 m\AA.
}\label{fig:std} \vspace{-0.9cm}
\end{figure}

\section{Origin of the dark lanes}

To find the origin of these dark lanes in FWHM images, we have
performed spectral line synthesis of the \FeI\ 5250.2 \AA\ line in
a snapshot of magneto-convection simulations \citep{Vogler2005}.
We considered a simulation run with a bi-polar magnetic field
structure and an average unsigned magnetic field in the box of 30
G, corresponding to a quiet solar region \citep[for more details
on this simulation snapshot see][]{Khomenko+etal2005}. The grid
size of the simulations is about 20 km. The synthesis was done in
LTE using the SIR code \citep{RuizCobo+delToroIniesta1992}. No
additional line broadening mechanisms, such as macro- or
microturbulence were introduced.

We found that the areas with extremely narrow intensity profiles
adjacent to intergranular lanes also appear in the simulations.
For an adequate comparison with the observations we have degraded
the simulation spectra in the following way. Firstly, we performed
a convolution of each monochromatic image with an Airy function
imitating the action of a 100 cm telescope. Secondly, we convolved
the spectra in wavelength with the transmission profiles of the
IMaX etalon. Finally, we fit a Gaussian function to the five
wavelength points and calculated the width of the Gaussian
profiles as a proxy to the intensity profile width, as well
as velocity, similar to the observations. The resulting image of
the simulated FWHM is shown in Figure \ref{fig:fwhm_sims}a.

The FWHM image shows structures similar to the observed ones, i.e.
larger FWHM in intergranular lanes and very low FWHM in the narrow
regions adjacent to lanes. Note that the average value and the
range of variations agree very well in the simulations and
observations.
Similar to the observations, the dark lanes in the FWHM image
correspond to the normalized continuum intensity around unity,
i.e. they mark the transition between granules and intergranular
lanes (Fig.~\ref{fig:fwhm_sims}b). Their width is a few pixels
($\sim$0\farcs2).

\begin{figure*}
\center
\includegraphics[width=17cm]{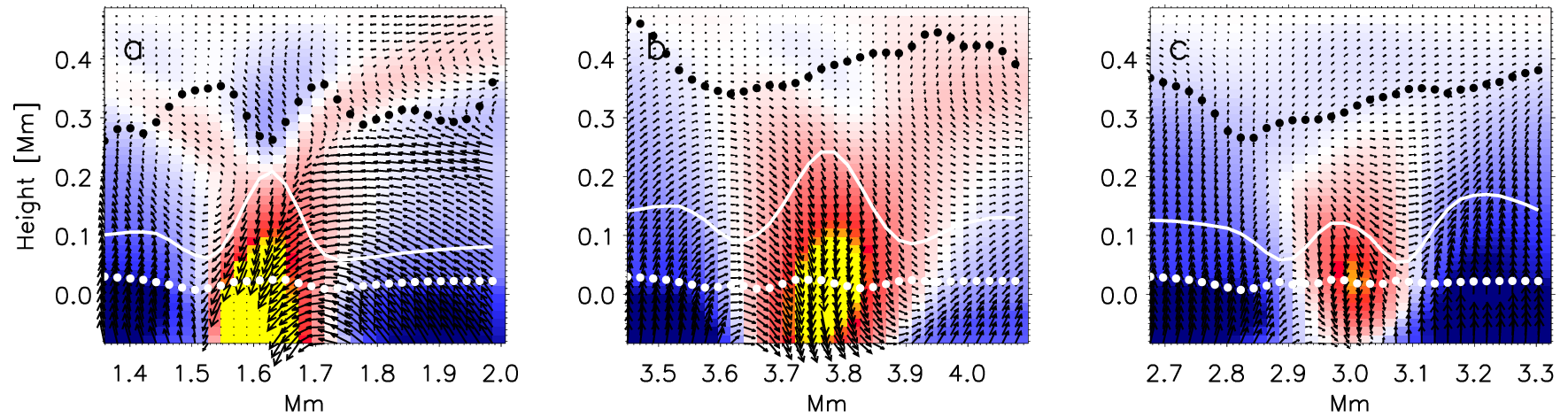}
\caption{Examples of the velocity field in the simulations at
locations marked (A), (B) and (C) in Fig.~\ref{fig:fwhm_sims}b.
The color coding indicates vertical velocity, with red being a
downflow and blue being an upflow, both saturated at 3 \kms.
The white and black dotted lines indicate the effective formation
height of the continuum and line core of the \FeI\ 5250.2 \AA,
respectively, estimated from the calculation of the Response
Functions. The white solid line shows horizontal variations of
FWHM in arbitrary units. }\label{fig:flows}
\end{figure*}

Due to the small width of these dark lanes, about 0\farcs2, high
resolution and image stability are required for their detection.
Figure \ref{fig:fwhm_sims}c shows the FWHM after the simulated
spectra were degraded to Hinode resolution and pixel size (note
that the \FeI\ 6301.5 \AA\ is used in this case). This image
preserves brightenings in intergranular lanes corresponding to
large FWHM, but the adjacent dark lanes do not show up at this
resolution. On the one hand, this comparison demonstrates that
IMaX/{\sc Sunrise} data quality is necessary to detect these
features. On the other hand, it proves that the instrument reached
indeed very high spatial resolution \citep{Berkefeld+etal2010}.

The IMaX spectral line of  \FeI\ 5250.2 at \AA\ is sensitive to
the velocity broadening, but also to magnetic fields via the
Zeeman effect, and to temperature. To find out which effect is
primarily responsible for the spatial variations of the FWHM, we
performed additional calculations. We repeated the spectral
synthesis for three cases: (1) magnetic field was set to zero,
preserving the variations in the rest of the variables; (2)
horizontal variations of the temperature and gas pressure were set
to zero using for all the points the mean stratification, but
preserving variations of the velocity and magnetic field;
(3) velocity was set to zero at all points, while preserving
variations in the rest of the variables.

We found that the magnetic field's influence on the horizontal
variations of the FWHM is negligible except for a few isolated
locations in intergranular lanes where the field reaches kG
values. There the FWHM is additionally enhanced by the Zeeman
broadening. Figure \ref{fig:std} gives the variations of the FWHM
in the cases (2) and (3). Case (2), where only velocity variations
are preserved, looks almost identical to the original case in
Fig.~\ref{fig:fwhm_sims}a. While case (3), where only variations
of the thermodynamic variables are preserved, looks very
different, showing almost an opposite behaviour to the case (2).
Note that, curiously, in the case (3) the broadening due to
magnetic field becomes evident in some intergranular lanes where
the field is strongest (locations at $X =1-2$ Mm; $Y=1-2$ Mm and
$X =4$ Mm; $Y=3$ Mm in the right panel of Fig.\ref{fig:std}). The
broadening due to magnetic field is much weaker compared to the
velocity broadening, but is comparable in magnitude to the thermal
broadening.
The average value of the FWHM and its standard deviation is much
lower for the case (3) as compared to the case (2) and to the
observations. Thus, we conclude that velocity broadening is the
prime responsible for the horizontal variations of the FWHM.

Figure~\ref{fig:flows} shows several examples of the velocity
field at the locations marked by letters in
Fig.~\ref{fig:fwhm_sims}b. At these locations the red line crosses
an intergranular lane with enhanced FWHM as well as the transition
between granule and intergranule with very low FWHM. As follows
from Fig.~\ref{fig:flows}, all these locations have features in
common. At the center of an intergranular lane, in the deep
photospheric layers, the vertical velocity is mainly a downflow
with a magnitude reaching about $3-4$ \kms. The intergranular
lanes are also characterized by a strong vertical velocity
gradient. In a few hundred kilometers, the velocity decreases from
$3-4$ \kms\ to zero, following the change of its sign at the top
of the atmosphere. These strong velocity gradients have immediate
implications on the width of the \FeI\ 5250.2 \AA\ line profiles,
whose core forms around 400 km \citep{Gurtovenko+Kostik1989}. The
action of these vertical velocity gradients over the height range
of formation of the line is similar to the action of
microturbulence velocity, producing additional broadening of the
profiles. We conclude that the large FWHM in intergranular lanes
is produced by a microturbulence-like effect due to strong
vertical velocity gradients, unlike the mechanism based on
super-sonic flows proposed by \citet{Nesis+etal1992}. Similar
conclusion, as for the enhancement of the FWHM in intergranular
lanes, was previously reached in the work of
\citet{Gadun+etal1997} from studying the line-parameter variations
in the two-dimensional hydrodynamic simulations.

Figure~\ref{fig:flows} shows that the regions with extremely low
FWHM correspond to the locations where the granular flow bends.
These locations are the transitions from granules with upflowing
velocities to intergranular lanes with downflows. As the velocity
vector bends there, the vertical velocity component is close to
zero and the vertical velocity gradient is very small as well. It
makes the velocity broadening at these transition locations
extremely low, thus producing narrow profiles, as detected in the
IMaX observations. In Fig.~\ref{fig:fwhm_sims}b the red
dashed contours mark the location with zero vertical velocity in
simulations. It shows that almost always these locations coincide
with dark lanes in FWHM image.

The IMaX data only have five points in wavelength, making it
impossible to check the conclusions about the velocity gradients,
obtained from simulations. However, the absolute values of the LOS
velocities are reliably obtained from these data. According to
simulations, the regions with extremely low FWHM correspond to
locations with almost zero vertical velocities. In
Fig.~\ref{fig:time} (right panel), we plot the LOS velocity
histograms obtained from the IMaX data over the whole time series.
This figure shows the existence of the three distinct families of
features. Granules, selected as locations with $I_c /\bar{I_c}>1$,
have the velocity histogram peak at $-0.6$ \kms. Intergranular
lanes, selected as locations with $I_c /\bar{I_c}<1$, show
downflowing velocities with the histogram peak at 0.4 \kms. This
histogram almost coincides with the one obtained by selecting
locations with enhanced FWHM given by white regions in
Fig.~\ref{fig:fwhm}b. Finally, there is a third distribution
corresponding to the borders between granules and intergranular
lanes. It is obtained by selecting locations with low FWHM given
by the black regions in Fig.~\ref{fig:fwhm}b. Its maximum appears
at zero LOS velocity. This observational confirmation supports the
conclusion based on simulations that the regions with extremely
small FWHM correspond to locations where the velocity vector
bends, producing a zero LOS velocity component.

\section{Conclusions}

In this Letter we have investigated the spatial distribution of
the full width at half maximum of the \FeI\ 5250.2 \AA\ intensity
profiles in the IMaX/{\sc Sunrise} data. We have shown that the
FWHM becomes large in intergranular lanes and is very small in
narrow 0\farcs2 wide strips adjacent to the lanes. The lifetime of
these narrow regions is the same as the intergranular lanes they
are attached to and they evolve in parallel. Based on the
similarity between IMaX observations and simulations of
magneto-convection, we have offered an explanation for the
formation of these profiles with small FWHM. We suggest that these
profiles appear in regions where the granular flow bends. There,
the velocity broadening is extremely small, as the LOS velocity
component is close to zero and has very small gradients. This
conclusion is supported by the analysis of the LOS velocity
histograms in the IMaX data. We claim that the {\sc Sunrise}/IMaX
resolution and image stability is the necessary condition to
detect these narrow transition zones between granules and
intergranular lanes.

\acknowledgments The German contribution to {\sc Sunrise} is
funded by the Bundesministerium f\"{u}r Wirtschaft und Technologie
through Deutsches Zentrum f\"{u}r Luft- und Raumfahrt e.V. (DLR),
Grant No. 50~OU~0401, and by the Innovationsfond of the President
of the Max Planck Society (MPG). The Spanish contribution has been
funded by the Spanish MICINN under projects ESP2006-13030-C06 and
AYA2009-14105-C06 (including European FEDER funds). The HAO
contribution was partly funded through NASA grant number
NNX08AH38G. This work has been partially supported by WCU grant
No. R31-10016 funded by the Korean Ministry of Education, Science,
and Technology.


\end{document}